\documentclass{article}
\usepackage{epsf,amssymb}

\newcommand{\be}{\begin{equation}}
\newcommand{\ee}{\end{equation}}
\newcommand{\bea}{\begin{eqnarray}}
\newcommand{\eea}{\end{eqnarray}}
\newcommand{\beas}{\begin{eqnarray*}}
\newcommand{\eeas}{\end{eqnarray*}}
\newtheorem{thm}{Theorem}

\def\One{\mathbb{I}}
\def\verttl{\;\raisebox{-4mm}{\epsfysize=10mm\epsfbox{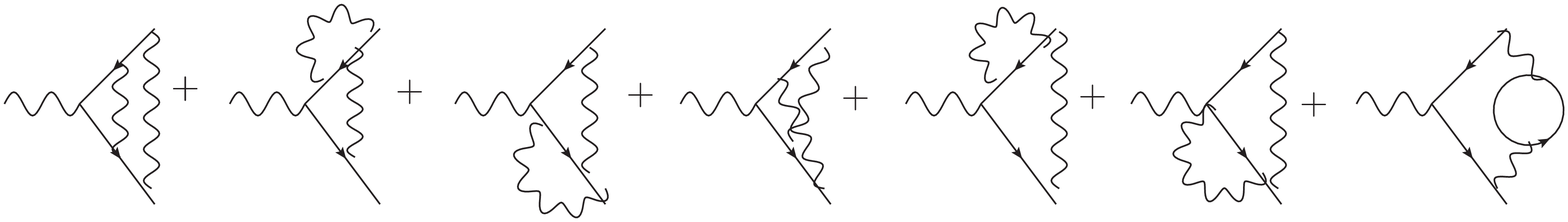}}\;}
\def\phol{\;\raisebox{-1mm}{\epsfysize=4mm\epsfbox{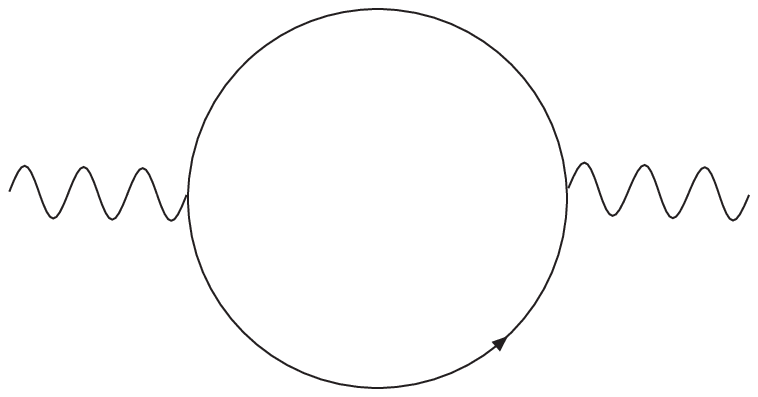}}\;}
\def\feol{\;\raisebox{-1mm}{\epsfysize=4mm\epsfbox{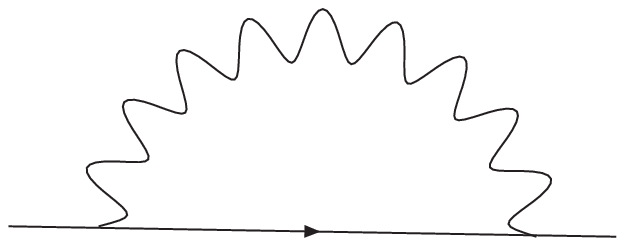}}\;}
\def\vertol{\;\raisebox{-1mm}{\epsfysize=4mm\epsfbox{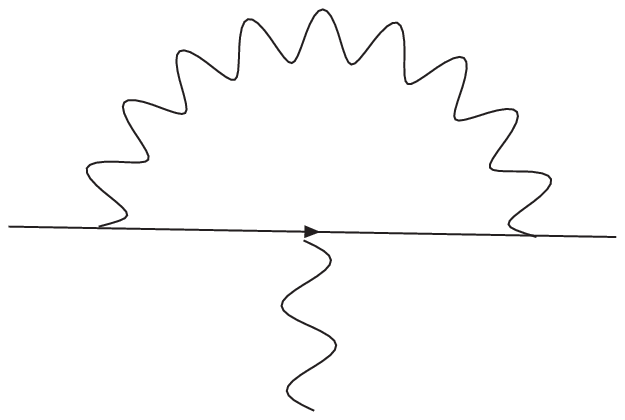}}\;}

\begin{document}

\title{Three \'Etudes in QFT}

\author{Dirk Kreimer
\thanks{CNRS, author also supported in parts by grant NSF/DMS-0603781 at Boston University, http://math.bu.edu/people/dkreimer.}\\
Institut des Hautes \'Etudes Scientifiques\\
91440 Bures sur Yvette, France\\
$^*$E-mail: kreimer@ihes.fr}

\maketitle

\begin{abstract}
We review QFT from the algebraic structure of its perturbative expansion and point out three new results.
\end{abstract}

%\keywords{Hopf algebras, perturbative QFT}

%\bodymatter

%\bodymatter

\section*{Acknowledgments}
It is a pleasure to thank Spencer Bloch for questions which ultimately led to Theorem \ref{kine} below.
The following is a write-up of an  invited contribution to the International Congress for Mathematical Physics (Prague, 2009).
Thanks are due to Pavel Exner and the other organizers.
\section{Algebraic structure in local QFT}
In recent years we have collected a considerable amount of algebraic structure underlying
the computational practice of quantum field theory, clarifying the mathematical foundations of
these computations and coming to new insights in QFT, see  \cite{KreimerAlg,BBK,grav,KWvS,vBKUY} and references there 
for more detail and notation. In this section, we give a succinct summary.
\subsection{Hopf and Lie algebras of graphs} There is a  Hopf and Lie algebras coming with 1PI Feynman graphs:
\be \Delta_{2n}(\Gamma)=\Gamma\otimes \One+\One\otimes\Gamma+\sum_{[\gamma]_{2n}}\gamma\otimes\Gamma/\gamma.\ee 
The sum is over all proper subsets  $[\gamma]_{2n}\subset\Gamma$ which constitute disjoint unions $\gamma=\prod\gamma_i$ of 1PI
graphs 
such that each $\gamma_i$ fulfills $\omega_{2n}(\gamma_i)\geq 0$. Here, $\omega_{2n}(\Gamma)
=2n|\Gamma|-\sum_{\mathrm{weights}\;w}w(\Gamma)$, and we sum over edge and vertex weights $w$.

These Hopf algebras include the Hopf algebra of renormalization for a theory renormalizable at $2n=D$ dimensions 
which then gives the forest formula  of renormalization theory.
With an antipode $S(\Gamma)=-\Gamma-\sum S(\gamma)\Gamma/\gamma$ we obtain renormalized Feynman rules $\Phi^R$
and counterterms $S_R$ as \be \Phi^R=m(S_R\otimes \Phi)\Delta,\,S_R=-R(S_R\otimes\Phi P)\Delta,\ee
from unrenormalized Feynman rules $\Phi$ and a projection into the augmentation ideal $P:H\to\mathrm{Aug}$.

For all these Hopf algebras there  is an underlying pre-Lie algebra structure:
\be [Z_{\Gamma_1},Z_{\Gamma_2}]=Z_{\Gamma_1}\otimes Z_{\Gamma_2}-Z_{\Gamma_2}\otimes Z_{\Gamma_1}\ee
with 
\be [Z_{\Gamma_1},Z_{\Gamma_2}]=Z_{\Gamma_2\star\Gamma_1-\Gamma_1\star\Gamma_2}.\ee
Here, $\Gamma_i\star\Gamma_j$ sums over all ways of gluing $\Gamma_j$ into $\Gamma_i$,
which can be written as
\be \Gamma_i\star\Gamma_j=\sum_\Gamma n(\Gamma_i,\Gamma_j,\Gamma)\Gamma.\ee
For any $\Gamma\in H$, we have a pairing
\be \langle [Z_{\Gamma_1},Z_{\Gamma_2}],\Gamma\rangle
=\langle Z_{\Gamma_1}\otimes Z_{\Gamma_2}-Z_{\Gamma_2}\otimes Z_{\Gamma_1},\Delta(\Gamma)\rangle,\ee
for consistency. With such section coefficients $n(\Gamma_i,\Gamma_j,\Gamma)$ we have
\be \Delta(\Gamma)=\sum_{h,g}n(h,g,\Gamma)g\otimes h.\ee
The sum is over all graphs $h$ including the empty graph and over all monomials in graphs $g$.

Here is an example of the coproduct $\Delta=\Delta_4$ on the sum of the two-loop vertex graphs $c_2^{\bar{\psi}A\!\!\!/\psi}$:
\beas \Delta\left(\verttl\right) & = & 3\vertol\otimes\vertol\\  +2\feol\otimes\vertol+\phol\otimes\vertol. & & \eeas
which reveals a sub-Hopf algebra structure.
\subsection{sub-Hopf algebras}
Along with such Hopf algebras comes the 
corresponding Hochschild cohomology and the sub-Hopf algebras $H_{\mathrm{grad}}$ generated by the grading and the quantum equations of motion.
If we let $c_k^r$ be the sum of all graphs contributing to an amplitude $r$ at $k$ loops
\be c_k^r=\sum_{\mathbf{res}(\Gamma)=r,|\Gamma|=k}\frac{1}{\mathrm{Aut}(\Gamma)}\Gamma,\ee
then $\Delta$ gives to these generators a sub-Hopf algebra structure:
Let $\mathcal{R}$ be the set of amplitudes which need renormalization, and let $H_{\mathrm{grad}}$ 
be the Hopf algebra spanned by generators
$\{\One,c_k^r\}$, $k\in\mathbb{N}$, $r\in\mathcal{R}$. The $c_k^r$ form a sub-Hopf algebra:
\be \Delta(c_k^r)=\sum_{j=0}^{k} P_{j,k}^s\otimes c_{k-j}^r,\ee
with $P^s_{j,k}$ a polynomial in generators $c_m^t$ of degree $j$. These sub-Hopf then allow for multiplicative renormalization
at a Lagrangian level.\\ 
\subsection{Symmetry} Co-ideals in the Hopf algebra of perturbation theory correspond
to symmetries in the Lagrangian:
Internal symmetries are in accordance with the Hochschild cohomology sructure of these Hopf algebras. 
For example the Ward identity in quantum electrodynamivs is equivalent to the statement that renormalized Feynman rules
can be defined on the quotient $H_{\mathrm{grad}}/I$, where $I$ is the ideal and co-ideal given by 
$c_k^{\bar{\psi}A\!\!\!\psi}+c_k^{\bar{\psi}\psi}=0.$ See \cite{KWvS} for further references, in particular for Walter van Suijlekom's work
on co-ideals as a reflection of internal symmetry.
\subsection{The renormalization group}
The coradical filtration and Dynkin operators govern the renormalization group and leading log expansion:
Indeed, in a leading-log expansion, terms $\sim \ln^j s$ (for $s$ a suitable kinematical parameter)
are obtained by the evaluation of
\be \sigma_k=\frac{1}{k!}\Phi^R m^{k-1} [S\otimes Y]^k\Delta^{k-1},\ee a fact heavily used in the reduction of Dyson--Schwinger equations to ODEs, see \cite{vBKUY}.
\subsection{Full amplitudes}
There is a  semi-direct product structure between the cocommutative Hopf algebra $H_{\mathrm{ab}}$ of superficially convergent amplitudes
and the Hopf algebra $H$  of amplitudes in $\mathcal{R}$ neding renormalization,
which is very handy in the organization of those Dyson--Schwinger equations:
\be H_{\mathrm{full}}=H_{\mathrm{ab}}\times H.\ee 
\subsection{The core algebra}
Finally the core Hopf algebra based on $\Delta_{\infty}$ has co-ideals \cite{KWvS} leading to recursions \`a la 
Britto-Cachazo-Feng-Witten, 
showing that loops and legs speak to each other in many ways. At the same time it allows an identification of 
renormalization as a limiting mixed Hodge structure \cite{BK}.
\section{Three new results}
In this section we outline three new results with details for each of them to be given elsewhere.
\subsection{generalized Witt algebras}
A graded commutative Hopf algebra $H$ can be regarded as the dual of the universal enveloping 
algebra $U(L)$ of a Lie algebra  $L$.
What is the dual Lie algebra  $L_{\mathrm{grad}}$  such that 
$U(L_{\mathrm{grad}})=H_{\mathrm{grad}}$?
Again, we need
\be \langle z_m^r\otimes z_n^s-z_n^s\otimes z_m^r,\Delta c_j^t\rangle=\langle [z_n^s,z_m^r],\Delta c_j^t \rangle,\ee
$\forall j>0,t\in\mathcal{R}$.

This certainly implies $t\in\{r,s\}$. 
Looking first at $r=s$,we get immediately 
\be [z_k^s,z_l^s]\sim (l-k)z_{k+l}^s.\ee
For the general case we find
\be [z_k^s,z_l^t]=-Q(s)kz_{k+l}^s+Q(t)lz_{k+l}^t.\ee
Here, $Q(s)$ are integers determined by the invariant charges of the theory, for example in QED one finds
$Q(\bar{\psi}A\!\!\!\psi))=Q(\bar{\psi}\psi)=2,Q(\frac{1}{4}F^2)=1$.
The check of the Jacobi identities is then a straightforward computation.
\begin{thm} With the above, $\mathcal{L}$ becomes a Lie algebra which is the Milnor-Moore dual to the Hopf algebra above.\end{thm}
Can we identify this Lie algebra? To this end, let us consider generalized Witt algebras \cite{Witt}.
We take $\mathbb{Q}$ as the underlying field (any field of characteristic zero would be possible)
and let  $|\mathcal{R}|$ be the number of amplitudes needing renormalization.

For $s\in\mathcal{R}$, consider $D_s:=\partial/\partial_{x_s}$ and define 
\be W^+_{\mathcal{R}}:=\mathrm{Span}_{\mathbb{Q}}\{x^qD_s|q\in \mathbb{Z}_+^{|\mathcal{R}|},1\leq s\leq|\mathcal{R}|\}.\ee
Then, $W^+_{\mathcal{R}}$ is the derivation Lie algebra of the polynomial ring $\mathbb{Q}[x_1,\cdots,x_{|\mathcal{R}|}]$
which is a simple Lie algebra of infinite dimensions and a subalgebra of the generalized Witt algebra $W$.

For integers $Q(t)$ as above, set 
\be z_m^s:=\left[\prod_{t\in\mathcal{R}}x_t^{Q(t}\right]^mx_s\partial_{x_s}.\ee
This puts $L_{\mathrm{grad}}\subset W^+$. We have no space for a proper discussion of 
physics aspects here, but mention that we can now augment the algebra $W^+$ by an element $Y$ such that 
$[Y,z_1^q]=z_1^q$.
We then get with
\be r:=Y\otimes z_1^q-z_1^q\otimes Y\ee a R-matrix for $L_{\mathrm{grad}}$.
Indeed, one immediately checks that $r$ fulfills the Yang--Baxter equation.
As direct consequence, we have Drinfel'd twists and quantized universal enveloping algebras \cite{Witt} available for future study.

\subsection{A massive co-ideal}
The Ward identity of quantum electrodynamics is well known: it relates the derivative wrt $p_\mu$ of the inverse fermion propagator
to the three-point vertex of a spin-one photon coupling at zero momentum to the corresponding pair of fermions: $ \partial_{p_\mu}G^{\bar{\psi}\psi}(p^2/\mu^2,m^2)=G^{\bar{\psi}A\!\!\!\psi}(p^2/\mu^2,m^2)$.

The proof is entirely graphical, upon recognizing that we can route the external momentum $p$ through the open fermion line for each graph contributing to 
$G^{\bar{\psi}\psi}$, and the Leibniz rule ensures that now each internal free propagator obtains a marking corresponding to the zero-momentum coupling of an external spin-one boson. 
Upon recognizing that this is the sum over all vertex graphs contributing to the descibed vertex function, we have our Ward identity.

This can be repeated similarly for the derivative $\partial_m G^{\bar{\psi}\psi}=G^{\bar{\psi}\psi\phi}$ which relates the inverse fermion propagator
to the coupling at zero-momntum transfer with a spin-zero scalar.
\begin{thm} This identity corresponds to a co-ideal in the Hopf algebra 
\be c_k^{\bar{\psi}\psi}+c_k^{\bar{\psi}\psi\phi}.\ee \end{thm}
This is straightforward to prove, what is interesting here is the interplay of Hopf co-ideals in theories with scalar particle and the interference with the Hopf algebra structure of perturbative gravity \cite{grav}.
\subsection{Kinematics as cohomology}
Let us consider a Green function $G^r$ and let us identify a suitable variable $s$ in general position such that
\be G_R^r(g,s,\Theta)=1\pm s^{\omega_r}\gamma_j^r(\{g\},\Theta)\ln^j s,\ee where $\Theta$ stands for a set of generalized angles
$p_i\cdot p_j/s$.
We have
\be G_R^r=\Phi_{\ln s,\Theta}^R(X^r)\ee
for renormalized Feynman rules $\Phi_{\ln s,\Theta}^R$ and $X^r$ a series which is a formal fixpoint of the equation
\be X^r=\One\pm \sum_k g^k B_+^{r;k}(X^rQ^k).\ee
The following theorem answers how the Hochshild cohomology of the perturbation expansion relates to the variation of physical parameters.
We assume $R$ is a kinematical subtraction scheme which renormalizes by subtraction at a fixed kinematical point, for example on shell.
\begin{thm}\label{kine}
\be \Phi^R_{\ln s_1+\ln s_2,\Theta}=\phi^R_{\ln s_1,\Theta}\star\phi^R_{\ln s_2,\Theta},\ee
and
\be \Phi^R_{\ln s,\Theta^\prime}\circ B_+^{r,k}=\Phi^R_{\ln s,\Theta}\circ \left(B_+^{r,k}+L^{r,k}\right),\ee
where $L^{r,k}$ is a exact one co-cycle:
$L^{r,k}=b\phi^{r,k}$.
\end{thm}
$b$ is the Hochschild boundary operator, $b^2=0$, and $\phi^{r,k}:H\to \mathbb{C}$ are suitable functions.
Note that this identifies the equivalence class of one-cocycles in the Hochschild cohomology 
with the space of kinematical variations at fixed scale $s$.

\end{document}